\documentclass[conference]{IEEEtran}
\usepackage{fancyhdr}
\IEEEoverridecommandlockouts
\renewcommand{\IEEEauthorrefmark}[1]{%
  \raisebox{0pt}[0pt][0pt]{\textsuperscript{\footnotesize #1}}}
\usepackage{pgfplots}
\usepgfplotslibrary{groupplots}
\usepackage[ruled,vlined,linesnumbered]{algorithm2e}
\usepackage{booktabs}
\pgfplotsset{compat=1.18}
\title{FluidGPU: Fine-Grained Kernel Disaggregation for Large Model Inference on Heterogeneous GPUs \\}

\author{
\IEEEauthorblockN{
Tiancheng Hu\IEEEauthorrefmark{1}\textsuperscript{*},
Jin Qin\IEEEauthorrefmark{2}\IEEEauthorrefmark{3},
Lei Chen\IEEEauthorrefmark{2}\IEEEauthorrefmark{3},
Junhao Hu\IEEEauthorrefmark{1},
Yuzheng Wang\IEEEauthorrefmark{1}\textsuperscript{*},
Mingxing Zhang\IEEEauthorrefmark{4},
Chunwei Xia\IEEEauthorrefmark{5},\\
Yizhou Shan\IEEEauthorrefmark{\ddag},
Huimin Cui\IEEEauthorrefmark{2}\IEEEauthorrefmark{3},
Ting Cao\IEEEauthorrefmark{4},
Zheng Wang\IEEEauthorrefmark{5},
Tao Xie\IEEEauthorrefmark{1},
Chenxi Wang\IEEEauthorrefmark{2}\IEEEauthorrefmark{3}}
\IEEEauthorblockA{
\IEEEauthorrefmark{1}SCS, Peking University, Beijing, China}

\IEEEauthorblockA{
\IEEEauthorrefmark{2}SKLP, Institute of Computing Technology, CAS }
\IEEEauthorblockA{
\IEEEauthorrefmark{3}University of Chinese Academy of Sciences, Beijing, China}

\IEEEauthorblockA{
\IEEEauthorrefmark{4}Tsinghua University, Beijing, China}
\IEEEauthorblockA{
\IEEEauthorrefmark{5}University of Leeds, West Yorkshire, England}

\thanks{\textsuperscript{*} Part of the work was done during the internship at SKLP, Institute of Computing Technology, CAS.}
\thanks{\ddag\ Independent Researcher}
}

\newcommand{\name}[0]{FluidGPU\xspace}

\newcommand{\codeIn}[1]{{\small\texttt{#1}}}
\newcommand{\modify}[1]{\textcolor{black}{#1}}

\newcommand{\eg}{\hbox{\emph{e.g.}}\xspace}
\newcommand{\ie}{\hbox{\emph{i.e.}}\xspace}

\newcommand{\MyPara}[1]{\noindent\textbf{\textit{#1}}~}

\definecolor{ForestGreen}{RGB}{34,139,34}
\definecolor{DarkGreen}{rgb}{0.0, 0.5, 0.0} 

\usepackage{threeparttable}
\usepackage{pifont}
\usepackage{times,color,graphicx,amsmath,array,subcaption}
\usepackage{xurl}
\usepackage{xspace}
\usepackage{hyperref}
\usepackage{tablefootnote}
\usepackage{listings}
\usepackage{tikz}

\lstdefinestyle{mystyle}{
      language=C++,
        basicstyle=\ttfamily\scriptsize, 
        keywordstyle=\color{javapurple}\bfseries,
        stringstyle=\color{javared},
        commentstyle=\color{javadocblue},
        morecomment=[s][\color{javadocblue}]{/**}{*/},
        numbers=left,
        breaklines=true,
        numberstyle=\tiny\color{black},
        stepnumber=1,
        numbersep=5pt,
        tabsize=2,
        showspaces=false,
        showstringspaces=false,
        morekeywords={foreach, uint64_t, uint16_t},
        classoffset=0,
        xleftmargin=1.8em,
        escapeinside={(*@}{@*)},
        moredelim=*[is][\color{red}]{[[[}{]]]},
        captionpos=b,
}

\begin{document}

\pagestyle{empty}

\bstctlcite{IEEEtran:BSTcontrol}

\maketitle
\thispagestyle{fancy}
\lhead{}
\rhead{}
\chead{}
\lfoot{\footnotesize{
SC26, November 15-20, 2026, Chicago, Illinois, USA
\newline 979-8-3195-4789-7/26/\$31.00 \copyright 2026 IEEE}}
\rfoot{}
\cfoot{}
\renewcommand{\headrulewidth}{0pt}
\renewcommand{\footrulewidth}{0pt}
\begin{abstract}

Disaggregation assigns components of an AI workload to different GPU types, providing a way to utilize modern heterogeneous GPU clusters. However, existing solutions operate at a coarse granularity and are tightly coupled to specific model architectures, leaving substantial performance untapped.
This paper presents \name, the first kernel disaggregation system designed to improve the performance and cost efficiency of large-model inference on heterogeneous GPUs. Our key insight is that kernels within a single application exhibit diverse resource demands, making kernel-level disaggregation the most suitable granularity for aligning computation with hardware capabilities.
\name combines offline analysis with online adaptation. It extracts precise inter-kernel dependencies from PTX to ensure correctness, overlaps communication with computation through pipelined execution, and uses workload-aware scheduling with lightweight runtime adaptation. \modify{Our evaluation across five heterogeneous GPUs and four model architectures, scaling up to 16 GPUs, shows that \name achieves up to 2.3$\times$ the serving throughput of existing disaggregation approaches while generalizing to model architectures where prior approaches do not apply. Its cost efficiency, measured as throughput per dollar, is up to 1.6$\times$ that of these approaches. Our evaluation further shows that \name on a heterogeneous GPU pair can, in some cases, deliver more than twice the throughput of the high-end GPU alone while costing less than two high-end GPUs.}
\end{abstract}

\section{Introduction}\label{sec:intro}
Modern GPU data centers are increasingly heterogeneous, integrating a wide range of GPU types and generations. This heterogeneity stems from the mismatch between GPU release cycles and retirement schedules, driven by high costs and limited supply~\cite{ppipe, tco_problem}. For example, Google Cloud offers both high-end GPUs (\eg, A100 and H100) and general-purpose GPUs (\eg, L4 and RTX Pro 6000)~\cite{google_cloud}, and AWS shows a similar trend~\modify{\cite{AWS_types}}. At the same time, the rapid growth of AI has dramatically increased demand for GPU resources~\cite{lammertyn2024chatgpt}. This demand is further amplified by the emergence of agentic AI applications, where a single user request may trigger multiple sequential model invocations, significantly increasing inference workload and cost~\cite{cost_dynamic_reasoning, agentix}.

In this context, a primary challenge for cloud providers is maximizing the serving performance of heterogeneous GPU clusters while improving cost efficiency (Perf/\$)~\cite{cost-efficiency1, splitwise}. However, existing scheduling mechanisms fail to fully exploit the architectural diversity of heterogeneous GPUs, leaving substantial performance and cost-efficiency gains untapped.

\MyPara{Existing approaches.}
Recent studies \cite{distserve, cronus, megascale-infer, step-3, splitwise} \modify{use disaggregation to exploit GPU heterogeneity}. These approaches partition AI workloads according to their computational characteristics and route them to the most suitable GPU. This strategy is particularly common in Transformer-based LLM inference, where distinct phases or blocks have different resource demands. For example, prefill-decode (PD) disaggregation assigns the compute-intensive prefill phase to high-TFLOPS GPUs, while offloading the memory-bound decode phase to GPUs with higher memory bandwidth \cite{hexgen}. Similarly, block-level disaggregation schemes, such as Attention-FFN disaggregation \cite{megascale-infer}, aim to better align computation with hardware specialization.

Despite their effectiveness, existing disaggregation approaches suffer from two fundamental limitations. First, they operate at a coarse granularity (\eg, phases or computation blocks), leaving substantial performance untapped. For example, even within a single FFN block, different kernels may exhibit markedly different resource demands, which coarse-grained disaggregation fails to capture. Second, these approaches are often application-specific and tightly coupled to particular model architectures, limiting their generality across model types.

\begin{figure}[t!]
    \centering
    \includegraphics[width=0.9\linewidth]{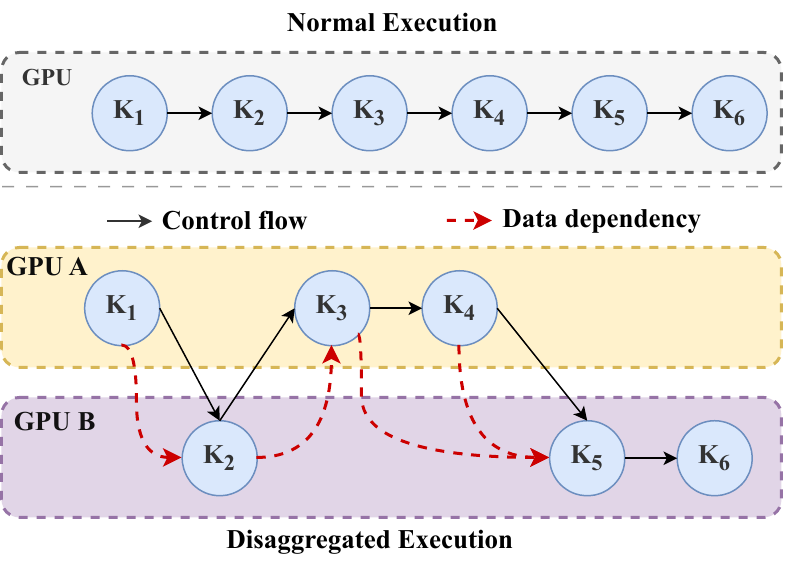}
    \caption{Simplified example of kernel disaggregation.}
    \label{fig:kernel_example}
\end{figure}

\MyPara{Key insight.}
We observe that the effective utilization of heterogeneous GPUs requires aligning \emph{hardware heterogeneity with application heterogeneity} at \emph{the right granularity}. On the hardware side, modern clusters consist of diverse GPUs with complementary strengths, including differences in compute capacity, memory bandwidth, and cost efficiency (\S\textcolor{DarkGreen}{\ref{sec:GPU heterogeneity}}). On the application side, the GPU kernel is the basic user-level scheduling unit and therefore a natural granularity for disaggregation. Within a kernel, finer granularities, such as thread blocks, are largely homogeneous and do not benefit from further partitioning. In contrast, coarser granularities, such as execution phases, fail to capture important variations in resource demand (\S\textcolor{DarkGreen}{\ref{sec:limitation}}). Moreover, kernels within a single application exhibit diverse \modify{resource behavior}, ranging from compute-bound to memory-bound, which indicates strong heterogeneity at the kernel level (\S\textcolor{DarkGreen}{\ref{sec:kernel heterogeneity}}).


\MyPara{Our approach.} 
We present \name, the first system to exploit heterogeneous GPUs through fine-grained kernel disaggregation. Given a set of heterogeneous GPU resources, \name aligns GPU heterogeneity with kernel heterogeneity to maximize overall performance and cost efficiency. Figure~\textcolor{DarkGreen}{\ref{fig:kernel_example}} shows a simplified example of kernel disaggregation. Realizing this approach requires addressing three key challenges.

The \textit{first challenge} is preserving functional correctness under kernel disaggregation. Distributing kernel execution across GPUs makes data dependencies that are implicit in local execution explicit. A kernel scheduled on one GPU may depend on data produced by a preceding kernel on another GPU. To prevent data hazards, the system must analyze the memory addresses and sizes of buffers accessed by each kernel. This analysis is difficult in modern AI frameworks, relying on opaque kernels such as user-defined or JIT-compiled kernels whose semantics are unavailable at compile time. We therefore design a \modify{PTX-level kernel analyzer}~\cite{ptx}. By instrumenting memory-access instructions, the analyzer captures precise buffer boundaries and extracts read-after-write (RAW) dependencies, thereby preserving correctness (\S\textcolor{DarkGreen}{\ref{sec:kernel analyzer}}).


The \textit{second challenge} is mitigating communication overhead and GPU underutilization. Kernel disaggregation introduces inter-GPU transfers for cross-device data dependencies. When dependent kernels are placed on different GPUs, downstream kernels must wait for upstream completion, leaving GPUs idle. To address this issue, \name employs a pipelined request-processing mechanism that executes multiple requests concurrently across heterogeneous GPUs. When a GPU stalls waiting for remote data, the GPU worker schedules ready kernels from other requests, overlapping communication with computation and maintaining high utilization (\S\textcolor{DarkGreen}{\ref{sec:GPU worker}}).

The \textit{third challenge} is designing an optimized scheduling policy. An unsuitable scheduling policy may incur excessive inter-GPU communication and poor load balance, causing system-level overheads to outweigh computation gains. To address this issue, \name derives workload-aware scheduling policies for different optimization objectives. For offline workloads, which prioritize system throughput, \name formulates scheduling as a mixed-integer linear programming (MILP) problem that jointly models kernel characteristics, compute throughput, communication overhead, and GPU load balance. For latency-sensitive online workloads (\eg, LLM serving), \name employs a latency-oriented scheduling policy (\S\textcolor{DarkGreen}{\ref{sec:policy_planner}}). A lightweight online monitor dynamically adjusts these policies as request loads change (\S\textcolor{DarkGreen}{\ref{sec:online monitor}}).

We integrate \name into popular vLLM \cite{vLLM} and PyTorch \cite{pytorch} frameworks, and evaluate \name across five heterogeneous GPUs and four model architectures, scaling up to 16 GPUs. \modify{Compared with existing disaggregation approaches, \name achieves up to 2.3$\times$ their serving throughput. Its cost efficiency, measured as throughput per dollar, reaches up to 1.6$\times$ that of these approaches.} Furthermore, \modify{\name sustains a 1.3$\times$ higher request rate under the same SLO}. \modify{Our evaluation further shows that \name on a heterogeneous GPU pair can, in some cases, deliver more than twice the throughput of the high-end GPU alone while costing less than two high-end GPUs.}
In summary, this paper presents the following main contributions:

\begin{itemize}
    \item We survey GPU architectural heterogeneity across modern data center GPUs and systematically study kernel-level heterogeneity across diverse AI workloads.

    \item We show that existing disaggregation approaches are too coarse-grained to capture kernel-level heterogeneity, fundamentally limiting their ability to exploit heterogeneous GPUs.

    \item We design and implement \name, the first kernel disaggregation system for heterogeneous GPUs, and release the source code at \hyperlink{0}{https://github.com/ICTPLSys/FluidGPU}.
    
    \item We conduct a comprehensive evaluation demonstrating that \name consistently outperforms state-of-the-art disaggregation approaches in both throughput and cost efficiency, while generalizing to model architectures where prior approaches are inapplicable.
\end{itemize}

\section{Background \& Motivation}

\begin{table}[]
\centering
\footnotesize
\begin{tabular}{cccccc}
\hline
                                                                & A100 & H100 & B200       & L40s  & RTX Pro 6000 \\ \hline
\begin{tabular}[c]{@{}c@{}}CUDA core \\ (TFLOPS)\end{tabular}   & 19.5 & 67   & 80         & 91.6  & 120          \\ \hline
\begin{tabular}[c]{@{}c@{}}Tensor core \\ (TFLOPS)\end{tabular} & 312  & 989  & $\sim$2500 & 366.5 & $\sim$500    \\ \hline
\begin{tabular}[c]{@{}c@{}}L2 cache \\  (MB)\end{tabular}       & 40   & 50   & 126        & 96    & 126          \\ \hline
Memory (GB)                                                     & 80   & 80   & 192        & 48    & 96           \\ \hline
Bw. (GB/s)                                                      & 1935 & 3350 & $\sim$8000 & 864   & 1597         \\ \hline
Price                                                           & 1.5  & 2.9  & 5.0        & 1     & 1.2          \\ \hline
\end{tabular}
\caption{Performance metrics and relative cost of different GPU architectures. \modify{CUDA-core and Tensor-core performance are reported in FP32 and BF16 TFLOPS, respectively.} The prices are derived mainly from Google Cloud's standard instance rates, normalized by the L40s \cite{google_price}.}
\label{tab:GPU heterogeneity}
\end{table}

\subsection{GPU Heterogeneity} \label{sec:GPU heterogeneity}

Modern GPU clusters are inherently heterogeneous. GPU models make different trade-offs across key architectural dimensions, including compute throughput, memory bandwidth, memory capacity, and cost efficiency.
As shown in Table~\ref{tab:GPU heterogeneity}, no single GPU dominates across all dimensions. \modify{For example, compared to the A100, the L40s delivers 4.7$\times$ higher CUDA-core throughput and 1.2$\times$ higher Tensor-core throughput, but provides only 45\% of the A100's memory bandwidth. Similarly, while the B200 offers near-leading performance across most hardware metrics, GPUs such as the L40s and RTX Pro 6000 provide up to 7.4$\times$ the compute throughput per dollar of high-end GPUs.}
These complementary strengths make heterogeneous GPU deployments both practical and economically attractive.

This heterogeneity is a sustained production trend rather than a transient condition. GPU vendors continuously release devices that make different trade-offs among multiple dimensions, so future deployments are also likely to remain heterogeneous rather than converging to a single dominant GPU type. This trend is further reinforced by platforms such as NVIDIA MGX~\cite{MGX}, which have been adopted by multiple vendors, including Supermicro~\cite{supermicro} and Gigabyte~\cite{gigabyte}, and explicitly support heterogeneous GPU configurations within a single server. Together, these trends motivate systems that can exploit heterogeneous GPUs effectively.

Comparable architectural diversity exists in other ecosystems, including AMD. Although broader heterogeneity across vendors or accelerator types (\eg, GPUs and NPUs) is also possible, it introduces additional challenges, including fragmented software stacks~\cite{SYCL} and the lack of a unified high-speed interconnect~\cite{lack_communication}. Given NVIDIA's dominant market position and mature software ecosystem, this paper focuses on heterogeneous GPUs within the NVIDIA platform.

\subsection{Intrinsic Kernel Heterogeneity} \label{sec:kernel heterogeneity}

\begin{figure}[t]
    \centering
    \includegraphics[width=\linewidth]{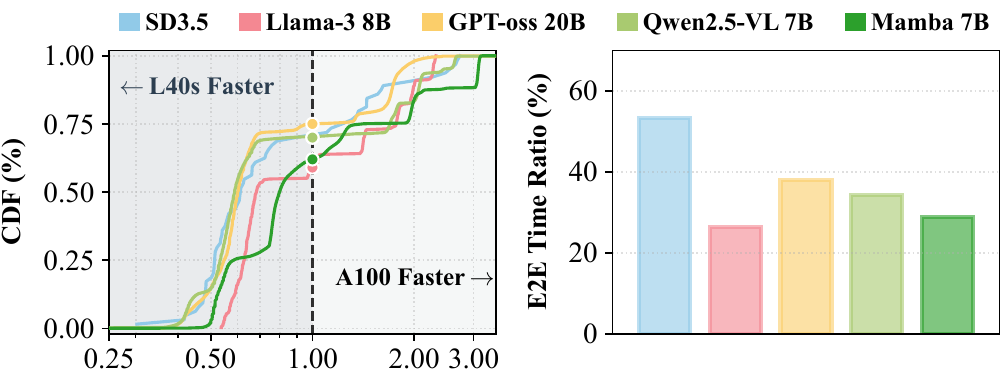}
    \caption{Kernel heterogeneity between A100 and L40s. (a) CDF of kernel execution-time ratios (L40s/A100) based on kernel count, and (b) share of A100 end-to-end (E2E) execution time attributable to L40s-faster kernels.}
    \label{fig:kernel heterogeneity}
\end{figure}

\name is designed to optimize GPU kernels that exhibit intrinsic heterogeneity in modern AI workloads. \modify{To quantify this heterogeneity, we profile five representative models spanning LLMs (Llama-3 8B and GPT-oss 20B), SSMs (Mamba 7B), MLLMs (Qwen2.5-VL 7B), and diffusion models (Stable Diffusion 3.5 (SD3.5)).} We measure individual-kernel execution time on two heterogeneous GPUs, A100 and L40s.

Figure~\textcolor{DarkGreen}{\ref{fig:kernel heterogeneity}} illustrates kernel heterogeneity from two perspectives. Figure~\textcolor{DarkGreen}{\ref{fig:kernel heterogeneity}(a)} shows the count-based CDF of kernel execution-time ratios (L40s/A100). On average, 67\% of kernels execute faster on L40s across models and tasks. However, kernel counts alone do not reflect their end-to-end impact. Therefore, Figure~\textcolor{DarkGreen}{\ref{fig:kernel heterogeneity}(b)} presents a time-weighted analysis, measuring the fraction of total A100 execution time contributed by kernels that run faster on L40s. These kernels account for 36\% of total execution time on average and up to 53\% for diffusion models.
These results show substantial opportunities to accelerate large-model inference on heterogeneous GPUs through fine-grained kernel disaggregation.

\subsection{Limitations of Coarse-Grained Disaggregation}
\label{sec:limitation}



\modify{Existing disaggregation approaches operate at either the phase level~\cite{distserve,splitwise} or the block level~\cite{megascale-infer}. Both granularities leave substantial performance untapped.}

\begin{figure}[]
    \centering
        \includegraphics[width=\linewidth]{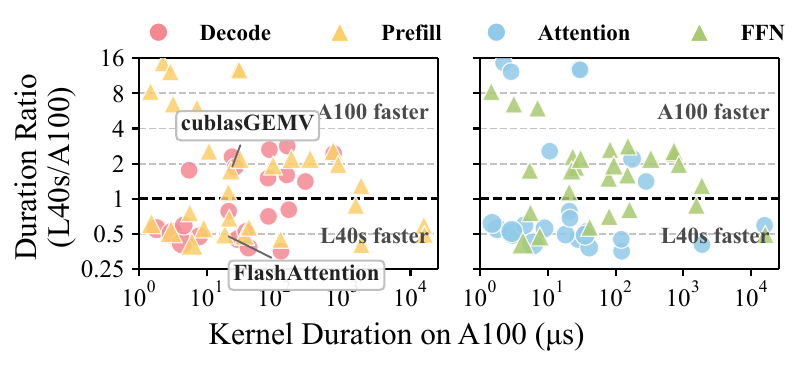}
        \caption{Kernel duration ratio (L40s\,/\,A100) for GPT-oss 20B inference, grouped by (a)~prefill and decode phases and (b)~attention and FFN blocks. Kernels above the dashed line run faster on A100, while those below run faster on L40s.}
    \label{fig:kernel_within_phase}
\end{figure}

Prefill-decode disaggregation~\cite{distserve,splitwise} assigns the entire prefill phase to one GPU and the decode phase to another. However, as shown in Figure~\textcolor{DarkGreen}{\ref{fig:kernel_within_phase}}(a), kernels within the same phase exhibit diverse GPU preferences: 45\% of prefill kernels and 57\% of decode kernels run faster on L40s than on A100. As a result, any phase-level disaggregation inevitably misplaces a large fraction of kernels.

To understand the root causes of kernel heterogeneity, we examine two representative prefill kernels. \texttt{cublasGemv}~\cite{cublas} is a widely used matrix–vector multiplication kernel that \modify{is strongly memory-bandwidth-bound}, with an operational intensity of approximately 1 FLOP/byte.
According to the roofline model~\cite{roofline}, its performance is primarily determined by memory bandwidth. A100's HBM provides approximately twice the bandwidth of L40s, yielding a 1.9$\times$ speedup.
In contrast, FlashAttention~\cite{flashattention} is compute-bound with high operational intensity that scales with sequence length. Its tile-based design maximizes data reuse in shared memory and the L2 cache, substantially reducing HBM traffic. As a result, performance is dominated by GPU Streaming Multiprocessor (SM)-level execution efficiency rather than memory bandwidth. Benefiting from a larger L2 cache, higher core frequency, and more efficient tensor cores, L40s achieves a speedup of up to 2.1$\times$ over A100.

Attention-FFN disaggregation~\cite{megascale-infer} operates at the block level, mapping attention and FFN to separate GPUs. However, as shown in Figure~\textcolor{DarkGreen}{\ref{fig:kernel_within_phase}}(b), kernels within the same block still exhibit divergent GPU preferences: 71\% of attention kernels run faster on L40s, and 33\% of FFN kernels run faster on L40s. Therefore, block-level disaggregation still misplaces a significant fraction of kernels, leading to suboptimal performance. These limitations motivate \name, which disaggregates at the kernel level to better exploit GPU heterogeneity.

\section{Design}

\name is the first system that disaggregates kernel execution across heterogeneous GPUs. Its primary design goal is correct and efficient kernel disaggregation.

\begin{figure}
    \centering
    \includegraphics[width=0.97\linewidth]{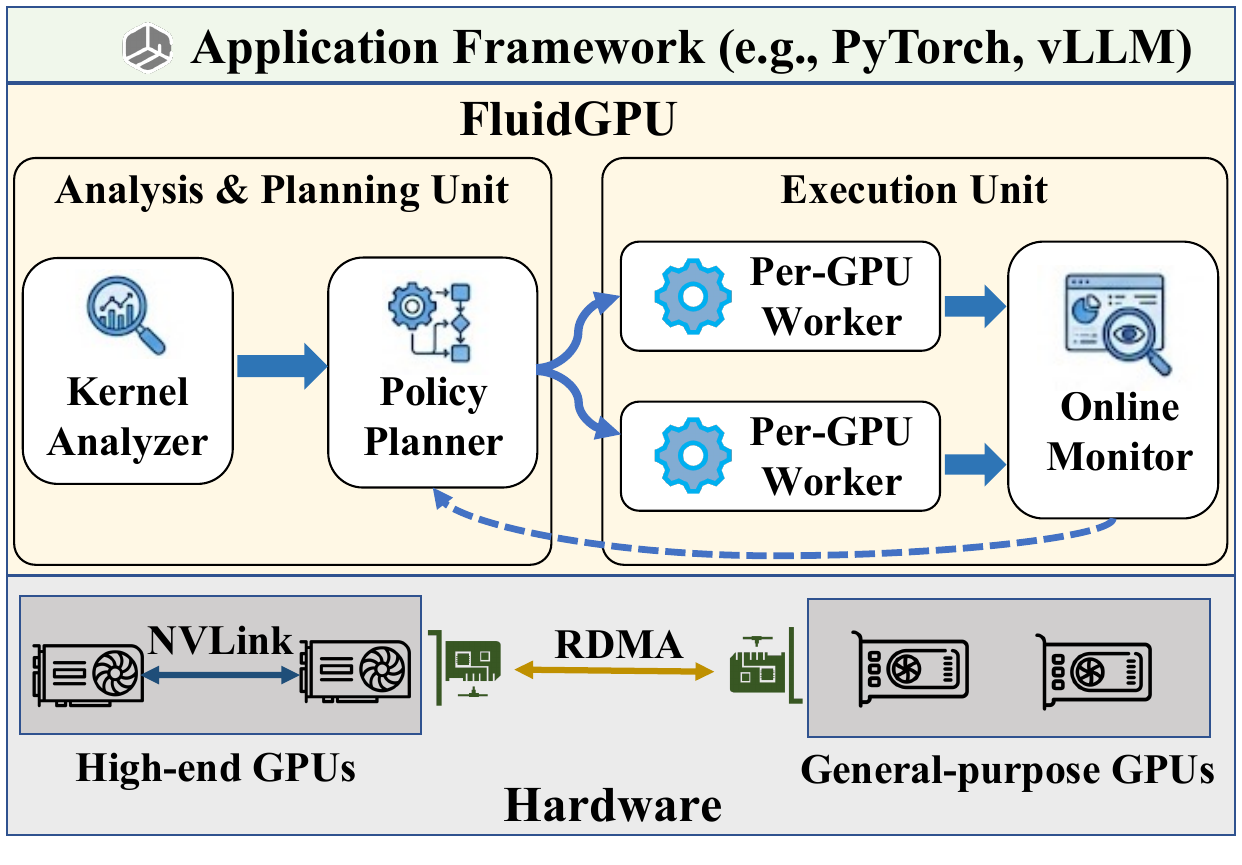}
    \caption{Design overview of \name.}
    \label{fig:overview}
\end{figure}

As shown in Figure~\textcolor{DarkGreen}{\ref{fig:overview}}, \name consists of four main components. The kernel analyzer profiles kernels offline to capture memory access patterns, extract inter-kernel data dependencies, and measure kernel latency on heterogeneous GPUs. Using these results, the policy planner derives workload-aware kernel scheduling policies that jointly consider kernel characteristics, communication overhead, and load balance. At runtime, the GPU worker \modify{executes the scheduling policy} for disaggregated kernel execution. Each worker executes only its assigned kernels and orchestrates the required inter-GPU communication. Meanwhile, the online monitor continuously collects runtime statistics (\eg, request and kernel latencies) to support dynamic policy adaptation under changing workloads.


\subsection{Kernel Analyzer}
\label{sec:kernel analyzer}




The kernel analyzer identifies accessed buffer addresses and sizes from PTX code to resolve inter-kernel data dependencies. In practice, GPU kernels fall into two categories. The first includes standard library kernels, such as \texttt{cublasSgemm}~\cite{cublas}, whose semantics and memory access patterns are well defined. The second comprises opaque computation kernels. \modify{Analyzing such kernels} is challenging, as they are often developer-defined or JIT-compiled, making their memory access patterns difficult to observe.

\MyPara{Memory access analysis.} 
For standard library kernels, identifying memory access patterns is straightforward because of their deterministic semantics. For example, the semantics of \texttt{cublasSgemm(.,A,B,.,C,.)} specify reads from \texttt{bufferA} and \texttt{bufferB} and a write to \texttt{bufferC}, while the API parameters provide the buffer sizes. In contrast, opaque kernels lack exposed semantics, making their memory access patterns difficult to infer.

PhoenixOS~\cite{phoenixos} represents prior work that identifies accessed buffers through GPU API argument speculation and runtime validation. Although this technique recovers buffer base addresses, it has a fundamental limitation. Modern AI frameworks such as PyTorch \cite{pytorch} employ internal memory managers (\eg, the caching allocator~\cite{cachingallocator}) that virtualize GPU memory independently of the CUDA runtime. Because CUDA allocation APIs are not invoked for every tensor creation, this technique often overestimates the actual buffer size and causes unnecessary communication in \name.

\name addresses this limitation with \modify{a kernel analyzer implemented at the PTX level}~\cite{ptx}. Following an approach similar to eBPF~\cite{NEUTRINO, eBPF}, \name instruments kernel code through PTX injection to track memory accesses. \modify{Listing~\textcolor{DarkGreen}{\ref{lst:ptx_instrumentation}} illustrates how instrumentation is performed on an opaque kernel.} Specifically, each global memory instruction (\eg, \texttt{st.global}, \texttt{ld.global}) is instrumented to record the virtual addresses that it accesses. Since the virtual addresses of buffers are contiguous, \name maintains only the minimum and maximum accessed addresses per instruction. These values are safely aggregated across concurrent threads using atomic operations (\eg, \texttt{atom.global}) to avoid data races. After kernel execution, \name retrieves the instruction-level $[min_k, max_k]$ intervals and aligns these aggregated intervals with the buffer base addresses to compute precise buffer sizes. The instruction opcode (\eg, \texttt{ld.global} or \texttt{st.global}) also indicates whether the identified buffer is read or written. Implementation details of the PTX analyzer are provided in \S\textcolor{DarkGreen}{\ref{sec:implementation}}.


\begin{lstlisting}[language=c++, 
caption={PTX instrumentation logic for an opaque kernel.},
float=t, 
moreemph={int, size_t, struct, void, __global__}, 
commentstyle=\color{ForestGreen}, % 设置注释为绿色,
emphstyle=\bfseries \color{blue},
morekeywords={(blockIdx.x + offset[0]), (blockIdx.y + offset[1]), offset[3], cudaLaunchKernel},
keywordstyle=\color{purple},
label={lst:ptx_instrumentation}]
.visible .entry opaque_kernel(
    ... // original parameters
    (*@\colorbox{gray!15}{.param  .u64    inst\_buf}@*)){
    .reg .b64   %rd<15>; // Declare registers
    .reg .b64   %unused
    ... // Load parameters
    (*@\colorbox{gray!15}{ld.param.u64       \%rd10, [inst\_buf];}@*)
    ... // Original kernel body
    (*@\colorbox{gray!15}{mad.wide.u32       \%rd11, k, 16, \%rd10;}@*)
    (*@\colorbox{gray!15}{atom.global.min.u64 \%unused, [\%rd11], \%rd1;}@*)
    (*@\colorbox{gray!15}{atom.global.max.u64 \%unused, [\%rd11+8], \%rd1;}@*)
    // Original k-th memory access instruction
    ld.global.f32 %f1, [%rd1]; 
    ... // Remaining kernel body
    }
\end{lstlisting}

\MyPara{Data dependency analysis.}
After determining the precise read and write semantics of each accessed buffer, \name constructs a \emph{data dependency graph} (DDG) that captures inter-kernel dependencies. Since kernel disaggregation preserves the original execution order, the analyzer needs to only identify true data dependencies, \ie, read-after-write (RAW) dependencies. Specifically, the analyzer maintains a global buffer registry that tracks the last writer of each buffer. When a kernel reads a buffer, the analyzer queries the registry to identify the most recent writer. If the writer differs from the current kernel, a dependency edge is added from the writer to the reader. By iterating over kernels in execution order, \name constructs a DDG where nodes represent kernels and directed edges encode RAW dependencies annotated with concrete buffer sizes. Asynchronous memory copies are treated uniformly as kernel nodes in the DDG, since their source, destination, and transfer size are explicitly specified in the API parameters.

The DDG captures intra-iteration dependencies by default. For buffers with cross-iteration RAW dependencies (\eg, KV caches in Transformer-based models~\cite{transformer}), the analyzer detects these dependencies by profiling multiple iterations and identifying buffers written in one iteration and read in subsequent iterations. \modify{The GPU workers handle these cross-iteration dependencies through per-GPU buffer replication and asynchronous delta transfers, as described in \S\textcolor{DarkGreen}{\ref{sec:GPU worker}}.}

Inference workloads may exhibit dynamic execution patterns, for example, due to varying tensor shapes or changes in computation driven by input sequence length. In practice, these patterns are not problematic for inference, as modern inference frameworks (\eg, vLLM~\cite{vLLM}, TensorRT \cite{tensorrt}) increasingly use CUDA Graph~\cite{cuda_graph} to reduce kernel launch overhead. Since CUDA Graph naturally requires a static execution pattern~\cite{static_execution_pattern}, \name can construct a separate DDG for each captured graph, capturing dynamic behaviors across different executions. For frameworks that do not yet support CUDA Graph capture, \name provides a preprocessing tool to automatically record the kernel execution trace and replay it under CUDA's stream capture API to produce an equivalent CUDA Graph.

\modify{Therefore, \name does not require the original application to use CUDA Graphs, but requires only a repeatable kernel sequence for each supported input shape. The profiling pipeline is executed once for each combination of hardware configuration and CUDA Graph or input-shape pattern. Subsequent requests matching the same pattern reuse the cached DDG and scheduling policy without repeating PTX instrumentation or dependency analysis. For data-dependent control flow, \name pre-generates a graph, DDG, and policy for each profiled path and falls back to single-GPU execution for unseen or non-capturable paths. Different requests therefore do not trigger continuous element-level dependency tracking. The offline preprocessing overhead is evaluated in \S\textcolor{DarkGreen}{\ref{sec:eval_ablation}}.}


In addition, the analyzer profiles the execution latency of each kernel on heterogeneous GPUs to inform scheduling-policy generation. \modify{The overall analysis is a one-time preprocessing step that can usually be completed during application warm-up.}




\subsection{Policy Planner}
\label{sec:policy_planner}
The policy planner generates a kernel-to-GPU scheduling policy for each DDG produced by the kernel analyzer. Because performance goals vary across scenarios (\eg, latency-critical online serving vs. throughput-driven offline processing), \name supports both throughput- and latency-oriented policies.

\MyPara{Throughput-oriented policy.}
This policy maximizes system throughput and targets offline inference workloads that seek to process as many requests as possible within a given time window (\eg, batched document processing). We formulate this mapping strategy as a mixed-integer linear programming (MILP) problem, with key terminology summarized in Table~\textcolor{DarkGreen}{\ref{tab:math_notations}}.

The solver first introduces a set of binary placement variables $x_{k,g}$, where $x_{k,g} = 1$ indicates that kernel $k$ is assigned to GPU $g$. Each kernel must be assigned to exactly one GPU:
\[
\sum_{g \in G} x_{k,g} = 1, \quad \forall k \in K.
\]

For each GPU $g \in G$, the computation time $T_g$ is the aggregated execution latency of all kernels assigned to it:
\[
T_g = \sum_{k \in K} t_{k,g}\, x_{k,g}
\]

The communication overhead $M_g$ captures the total transfer cost incurred by all incoming cross-GPU dependency edges targeting GPU $g$:
\[
M_g = \sum_{(i,j) \in E}\; \sum_{u \in G \setminus \{g\}} c_{ij}^{u,g} y_{ij}^{u,g},
\]
\[
c_{ij}^{u,g} = \ell_{u,g} + \frac{d_{ij}}{bw_{u,g}}
\]
Here, $y_{ij}^{u,g} \in \{0, 1\}$ encodes the joint placement of edge $(i,j)$, \ie, $y_{ij}^{u,g} = x_{i,u} x_{j,g}$. To keep the formulation strictly linear, \name applies standard boolean linearization techniques~\cite{integer} to replace this quadratic term.

Under pipelined asynchronous execution, each GPU alternates between computation and communication. The per-request stage time on GPU $g$ is therefore bounded by its slower component:
\[
W_g = \max(T_g,M_g)
\]
Following prior pipeline analysis~\cite{pipeDream}, the steady-state throughput of the entire system is determined by the slowest stage. 
Maximizing overall throughput is thus equivalent to minimizing the maximum stage time. The final objective is

\[
\min\; \max_{g \in G}\; W_g
\]

\begin{table}[]
\centering
\footnotesize
\caption{Terminology used in the problem formulation.}
\label{tab:math_notations}
\begin{tabular}{l p{0.68\linewidth}}
\hline
\textbf{Symbol} & \textbf{Description} \\
\hline
$G$ & Set of available heterogeneous GPUs. \\
$K$ & Set of kernels. \\
$E$ & Set of data-dependency edges in the DDG. \\
$t_{k,g}$ & Profiled latency of kernel $k$ on GPU $g$. \\
$d_{ij}$ & Size of the buffer transferred from kernel $i$ to $j$. \\
$bw_{u,g}$ & Communication bandwidth between GPU $u$ and $g$. \\
$\ell_{u,g}$ & Base communication latency between GPU $u$ and $g$. \\
$c_{ij}^{u,g}$ & Data transfer overhead for edge $(i, j)$ crossing from GPU $u$ to $g$. \\
$x_{k,g}$ & Binary variable equal to $1$ if kernel $k$ is assigned to GPU $g$ and $0$ otherwise. \\
\hline
\end{tabular}
\end{table}

\MyPara{Latency-oriented policy.}
The latency-oriented policy targets online serving scenarios where the objective is to minimize per-request end-to-end latency. At low load, limited request concurrency prevents \name from effectively overlapping computation and communication through pipelined execution. As a result, the per-request latency can be decomposed into two additive components: the total kernel execution time and the cumulative cross-GPU transfer overhead incurred along cut dependency edges.
\modify{Under such conditions}, the objective of the MILP solver is
$$ \min \left( \sum_{k \in K} \sum_{g \in G} t_{k,g} \cdot x_{k,g} + \sum_{(i,j) \in E} \sum_{u \in G} \sum_{g \in G \setminus \{u\}} c_{ij}^{u,g} y_{ij}^{u,g} \right) $$

The resulting MILPs can be solved efficiently with standard optimizers such as Gurobi~\cite{Gurobi}. The number of decision variables scales with $|K| \times |G|$, where $|K|$ denotes the number of kernels per DDG and $|G|$ the number of GPUs. In a typical configuration from our evaluation, GPT-oss 20B on an A100+L40s node has $|G|=2$ and $|K|\approx500$. Gurobi solves each DDG in about 20 ms, and enumerating all execution patterns takes roughly 5 seconds in total. \modify{This overhead is negligible because MILP optimization is performed offline and incurs no runtime overhead.} The scalability of the MILP solver is discussed in \S\textcolor{DarkGreen}{\ref{sec:eval_ablation}}.

\subsection{GPU Workers}
\label{sec:GPU worker}

At runtime, \name launches one GPU worker on each GPU, and each worker executes its assigned portion of the computation.

\MyPara{Disaggregated kernel execution.}
Based on the scheduling policy, each GPU worker executes only its assigned kernels. Because disaggregated execution requires explicit cross-GPU data transfers along cut dependency edges~\cite{stratacl}, \name adopts GPU-initiated communication via InfiniBand GPUDirect Async (IBGDA)~\cite{IBGDA}, as supported by NCCL~\cite{nccl}, to eliminate control-path overhead from CPU–GPU synchronization. This CPU-bypass mechanism allows \texttt{send} and \texttt{recv} operations to be issued as GPU-side kernels.

For a kernel $k$ with incoming cut dependency edges, the worker posts the corresponding \texttt{recv} kernels before launching $k$. After $k$ completes, if it has outgoing cut edges, the worker posts the corresponding \texttt{send} kernels. \modify{The two streams enable communication to overlap with independent computation, while lightweight CUDA events synchronize only direct dependencies.}

\begin{figure}
    \centering
    \includegraphics[width=0.85\linewidth]{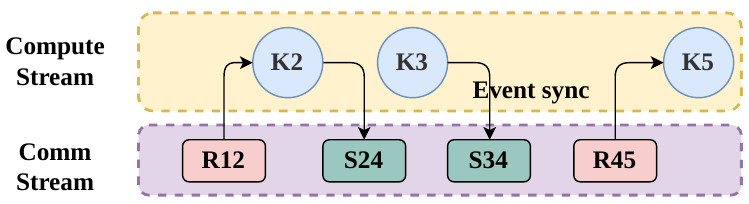}
    \caption{\modify{An execution example of the GPU worker. Compute kernels and communication operations run on separate streams and synchronize through CUDA events. $R_{ij}$ and $S_{ij}$ label receive and send operations using their corresponding DDG edge identifiers.}}
    \label{fig:comm}
\end{figure}

\modify{Figure~\textcolor{DarkGreen}{\ref{fig:comm}} illustrates dependency-local synchronization within one GPU worker. The receive operation $R_{12}$ must complete before $K_2$ starts. After $K_2$ completes, $S_{24}$ proceeds on the communication stream while the independent $K_3$ continues on the compute stream. The send operation $S_{34}$ waits for $K_3$ to complete, while $K_5$ waits for $R_{45}$. Only operations connected by an event synchronize, allowing unrelated computation and communication to overlap within the same request.}

\MyPara{Dynamic execution patterns.}
The GPU worker handles dynamic execution patterns through the CUDA Graphs described in \S\textcolor{DarkGreen}{\ref{sec:kernel analyzer}}. Since the kernel analyzer and policy planner have already constructed a separate DDG and scheduling policy for each CUDA Graph offline, the GPU worker only needs to select only the correct policy at runtime. Each worker maintains a dispatch table that maps each CUDA Graph handle to its pre-built scheduling policy. When a worker intercepts a CUDA Graph launch, it looks up and executes the corresponding policy. For low-overhead replay, \name decomposes each CUDA Graph into per-GPU subgraphs with \texttt{send} and \texttt{recv} nodes embedded alongside compute nodes (\S\textcolor{DarkGreen}{\ref{sec:implementation}}).

\MyPara{Pipelined request processing.}
\modify{Disaggregated kernel execution without pipelining leads to substantial GPU idle time during communication.} To address this issue, \name employs a pipelined request-processing mechanism that executes multiple independent requests concurrently across heterogeneous GPUs using multiple CUDA streams~\cite{multi_stream}. Each request is assigned to a dedicated compute stream. When one stream is blocked by \modify{a data transfer required by a dependency}, the GPU hardware scheduler dispatches ready kernels from other streams. \modify{This asynchronous execution overlaps computation with communication and reduces GPU idle time.}
However, when multiple requests execute concurrently without differentiation, they compete equally for Streaming Multiprocessor (SM) resources and tend to progress through the pipeline at similar rates. As a result, their communication phases may align, causing all streams to stall on data 
transfers simultaneously and leaving the GPU idle. 

\modify{To address this issue of being idle, \name maintains a fixed pool of priority-ordered CUDA compute streams whose size is bounded by the configured pipeline depth. A priority encodes only the relative arrival order of requests currently in flight. Earlier requests use higher-priority streams, biasing the GPU scheduler toward their ready kernels and staggering communication phases~\cite{hummingbird}. When a request completes, its stream and priority slot return to the pool. The runtime then recomputes the relative order of the remaining requests and updates their live streams through \texttt{cudaStreamSetAttribute} with \texttt{cudaStreamAttributePriority}. The updated priority applies to only work submitted after the call and does not affect previously submitted or executing kernels~\cite{multi_stream}. Later requests reuse the available streams and priority slots. The effectiveness of pipelined request processing is evaluated in \S\textcolor{DarkGreen}{\ref{sec:eval_ablation}}.}

\MyPara{\modify{KV-cache delta synchronization.}}
\modify{\name does not transfer the entire KV cache between GPUs. Instead, each GPU maintains a KV-cache replica. After every decoding step, the GPU worker uses the paged-KV block table and slot mapping maintained by vLLM~\cite{vLLM} to determine only the addresses and lengths of the newly generated K and V slices. It asynchronously propagates these KV-cache deltas to peer GPUs through a dedicated communication stream and a separate NCCL communicator~\cite{nccl}, while latency-critical intra-iteration transfers use a higher-priority traffic class~\cite{NCCL_bandwidth_conflict}. The DDG records the KV cache's cross-iteration RAW dependencies and the corresponding modified byte ranges. Before a kernel reads an updated KV-cache range in the next decoding step, the GPU worker waits for the associated delta-transfer completion event. In our measurements, asynchronously transferring this KV-cache delta adds less than 0.1\% to end-to-end decoding latency.}

\MyPara{\modify{Composability with model parallelism.}}
\modify{\name is orthogonal to model parallelism and composes naturally with existing parallelism strategies. It operates below the model-parallelism layer and leaves existing communication operations and topology unchanged. \name schedules only the compute kernels between consecutive communication steps within each heterogeneous GPU group. For tensor parallelism (TP), collectives remain on the original homogeneous TP group, and \name optimizes compute kernels within each TP rank's heterogeneous GPU group, as evaluated in Figure~\textcolor{DarkGreen}{\ref{fig:cluster_scale}}(b). For sequence parallelism (SP) and Ring Attention~\cite{ringattention}, sequence-shard exchanges and ring send and receive operations remain unchanged, while local attention and GEMM kernels between two communication steps can be disaggregated. For expert parallelism (EP) in MoE models~\cite{megascale-infer}, expert dispatch and combine all-to-alls remain fixed, while expert GEMMs are scheduled across the heterogeneous GPU group. Because input-dependent expert routing can produce different kernel graphs, each profiled routing pattern requires a separate DDG and policy. An unprofiled pattern conservatively retains the original expert placement.}

\subsection{Online Monitor}
\label{sec:online monitor}
The online monitor targets serving scenarios where request arrival rates fluctuate unpredictably over time. Unlike offline workloads with fixed load, online serving must handle dynamic queueing pressure, making \modify{any static policy} suboptimal. A latency-oriented policy processes requests without pipelining, which underutilizes GPU resources and incurs high queueing delay under heavy load. In contrast, a throughput-oriented policy becomes less effective under light load due to limited concurrency and may introduce additional latency from suboptimal scheduling.

To adapt to workload dynamics, the monitor tracks per-request latency, per-kernel-group latency, and communication time. Instead of profiling individual kernels, \modify{it measures the interval between consecutive communication operations. This interval contains a sequence of kernels, which we define as a \textit{kernel group}.} This design reduces monitoring overhead while retaining sufficient information for effective policy selection. Based on these signals, \name performs queueing-aware policy switching at fixed intervals of length $W$.

At each window boundary, the monitor computes the average request latency $\bar{L}_{\mathrm{req}}$ and the execution-only latency $\bar{L}_{\mathrm{exec}}$, which aggregates computation and communication time while excluding queueing delay. The ratio $\bar{L}_{\mathrm{req}} / \bar{L}_{\mathrm{exec}}$ serves as an indicator of queueing pressure. A low ratio implies negligible queueing delay and favors the latency-oriented policy, whereas a high ratio indicates that queueing dominates end-to-end latency, triggering a switch to the throughput-oriented policy to improve system processing capacity.

The monitor introduces two key hyperparameters, the time window $W$ and the queueing threshold $\beta$, which can be tuned according to workload characteristics. We further analyze their sensitivity and the runtime overhead of the monitor in \S\textcolor{DarkGreen}{\ref{sec:eval_ablation}}.

\section{Implementation}
\label{sec:implementation}

The current implementation of \name targets NVIDIA GPUs. It comprises approximately 3K lines of C++/CUDA code for the GPU worker runtime and 2K lines of Python code for the kernel analyzer, including PTX instrumentation and data dependency analysis. \name is integrated with vLLM~\cite{vLLM} and PyTorch~\cite{pytorch}.

\MyPara{PTX analyzer.}
Building on the probe engine from NEUTRINO~\cite{NEUTRINO}, we implement a PTX-level kernel analyzer. To track memory accesses, we extend the kernel parameter list with an instrumentation buffer that is loaded through inserted \codeIn{ld.param} instructions. We then inject \codeIn{atom.global} instructions around each global memory access (\eg, \codeIn{ld.global}, \codeIn{st.global}) to record the minimum and maximum accessed addresses per instruction. The modified PTX is recompiled into machine code with \codeIn{ptxas}~\cite{ptxas}.

A known limitation is that the analyzer cannot handle indirect buffer accesses. For example, if a kernel accesses GPU buffers via global variables not listed in the kernel launch arguments, the analyzer cannot obtain the base buffer address to calculate the buffer size. However, a comprehensive study~\cite{phoenixos} shows that the most popular ML frameworks (\eg, PyTorch, vLLM) do not contain such kernels. Only one kernel in the legacy Rodinia benchmark~\cite{rodinia} exhibits this pattern. In these rare cases, \name conservatively executes the kernel on the same GPU without disaggregation.

\MyPara{CUDA Graph adaptation.}
\modify{For an application-provided CUDA Graph, \name partitions the compute nodes into cached per-GPU subgraphs in topological order according to the scheduling policy. It inserts GPU-initiated \texttt{send} and \texttt{recv} operations as kernel nodes at cross-GPU dependency edges and represents CUDA event record and wait operations as event nodes. Each GPU worker replays its complete cached subgraph with a single \texttt{cudaGraphLaunch} rather than issuing individual kernels from the host. Compute kernels therefore remain back-to-back wherever no cross-GPU dependency intervenes, preserving the amortization of CPU-side launch and setup overhead provided by CUDA Graphs~\cite{graph_overhead}. Only communication required by a cut dependency separates compute regions, and the policy planner explicitly models this cost. A graph generated from a conventional execution trace follows the same partitioning, caching, and replay procedure.}

\MyPara{Memory management.}
The current MILP formulation does not explicitly model per-GPU memory-capacity constraints and replicates the full model weights on every GPU during initialization, similar to PD disaggregation~\cite{distserve}. In practice, full replication is conservative because each GPU accesses only a subset of weight buffers, and the remaining unused buffers could be reclaimed. For offline workloads whose scheduling policy remains fixed, selective buffer reclamation can safely free memory for larger batch sizes.

\section{Evaluation}
\subsection{Experimental Setup}
\label{sec:eval_setup}

\begin{figure*}
    \centering
    \includegraphics[width=0.98\linewidth]{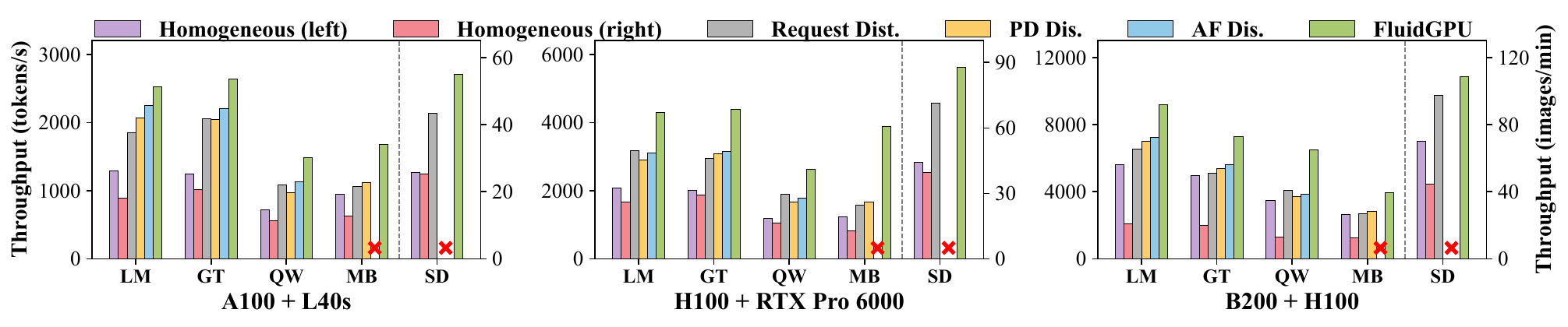}
    \caption{\modify{Offline throughput across five workloads and three heterogeneous GPU pairs. Request Dist. uses two independent replicas for every workload and GPU pair. The left y-axis shows tokens/s for LLM, MLLM, and SSM workloads, and the right y-axis shows images/min for SD. Red \textsf{X} marks indicate that a disaggregation approach is inapplicable.}}
    \label{fig:offline_throughput}
\end{figure*}   

\MyPara{Testbeds.}
\modify{We evaluate \name in two testbed environments.} Our main evaluation platform is a \textit{local setup} with three heterogeneous GPU pairs: 1 A100 + 1 L40s, 1 H100 + 1 RTX Pro 6000, and 1 B200 + 1 H100. Each GPU is equipped with a 200 Gbps RDMA NIC. We further evaluate \name's scalability in a \textit{cluster setup} with two distributed configurations. The first combines a node with 2 A100 GPUs and a node with 1 L40s GPU. The second combines a node with 8 B200 GPUs and a node with 8 H100 GPUs. Within each homogeneous high-end GPU group, GPUs are interconnected via NVLink, while cross-node communication is conducted over 200 and 400 Gbps RDMA NICs, respectively.

\MyPara{Workloads.}
To assess the generality of \name, we evaluate four model families:
(1) \textit{LLMs}: Llama-3 8B ($LM$)~\cite{Llama} and GPT-oss 20B ($GT$)~\cite{gptoss}, Transformer-based autoregressive models with distinct prefill and decode phases;
(2) \textit{SSMs}: Mamba-Codestral 7B ($MB$)~\cite{mamba_codestral}, which replaces attention with selective state-space layers for linear-time sequence generation;
(3) \textit{MLLMs}: Qwen2.5-VL 7B ($QW$)~\cite{qwen2.5-VL}, which augments an LLM decoder with a vision encoder for joint image-text reasoning;
and (4) \textit{Diffusion models}: Stable Diffusion 3.5 ($SD$)~\cite{sd3, sd35}, a text-to-image model based on the Multimodal Diffusion Transformer architecture, which synthesizes images through iterative denoising rather than autoregressive token generation. For \textit{offline evaluation}, the LLM and SSM workloads use the Splitwise conversational dataset~\cite{splitwise}, with a median input length of 1020 tokens and a median output length of 129 tokens. The MLLM workload uses images from the COCO captioning dataset~\cite{coco}, resized to 512$\times$512 pixels. The diffusion workload uses text prompts from the PartiPrompts dataset~\cite{parti} at 1024$\times$1024 resolution with 28 denoising steps. We measure the maximum sustainable throughput following common practice~\cite{nanoflow, helix}. For \textit{online evaluation}, we focus on GPT-oss 20B inference under both Poisson arrivals at varying request rates and a real-world trace from the Azure Conversation dataset~\cite{splitwise}, following previous work~\cite{hetis}.

\MyPara{Baselines.}
We compare \name with two state-of-the-art disaggregation approachs:

\begin{itemize}
\item \textbf{Prefill–decode disaggregation (PD Dis.):} a phase-level approach following DistServe~\cite{distserve}, which assigns the compute-intensive prefill phase and the memory-intensive decode phase to separate GPUs. \modify{This baseline does not apply because diffusion models lack a prefill–decode separation.}
\item \textbf{Attention–FFN disaggregation (AF Dis.):} a block-level approach following MegaScale-Infer~\cite{megascale-infer}, which maps memory-intensive attention and compute-intensive FFN operations to different GPUs. \modify{This baseline does not apply to SSM and diffusion models because these models do not expose the standard attention–FFN block structure assumed by AF disaggregation.}
\end{itemize}

\modify{We also include request distribution (Request Dist.) for offline throughput. Each GPU in the heterogeneous pair runs an independent vLLM replica, and incoming requests use the best-performing vLLM load-balancing policy among those evaluated. Unlike the disaggregation baselines, this scheme assigns each complete request to one GPU and performs no cross-GPU kernel execution.}

We also include homogeneous-GPU baselines without disaggregation. The inference engine is vLLM v0.18.0~\cite{vLLM}, with NCCL~\cite{nccl} as the communication backend. We enable standard scheduling techniques (\eg, continuous batching) and tune each disaggregation baseline to its best-performing GPU configuration for each hardware setting. To reduce latency jitter, we disable GPU dynamic voltage and frequency scaling (DVFS)~\cite{dvfs2}. Each experiment is repeated multiple times, with variance below 1\%, and we report the average results.



\subsection{End-to-end Performance}
\label{sec:eval_local}

\MyPara{Offline throughput.}
Figure~\textcolor{DarkGreen}{\ref{fig:offline_throughput}} reports the maximum sustainable throughput in offline serving across five workloads and three heterogeneous GPU pairs. We measure throughput as output tokens per second for LLMs, SSMs, and MLLMs, and as requests per minute for diffusion models. Across applicable configurations, \name outperforms PD disaggregation by 1.5$\times$ on average (up to 2.3$\times$) and AF disaggregation by 1.35$\times$ on average (up to 1.7$\times$).

\modify{Across the evaluated workloads and GPU pairs, Request Dist. falls below the ideal sum of the two single-GPU throughputs because of request-level load imbalance and routing overhead. For GPT-oss 20B on A100+L40s, its measured throughput is 2,052~tokens/s, whereas \name reaches 2,641~tokens/s, a 28.7\% improvement. Both configurations use the same request workload and both GPUs. This result shows that assigning whole requests to independent replicas cannot capture the within-request kernel preferences exploited by \name.}

The performance gains arise because \name captures kernel-level heterogeneity that coarser-grained approaches miss. PD disaggregation and AF disaggregation treat an entire phase or block as a single scheduling unit, forcing kernels with diverse resource demands onto the same GPU. In contrast, \name schedules kernels independently, aligning fine-grained kernel heterogeneity with hardware characteristics.

Existing disaggregation approachs also have limited applicability. \modify{AF disaggregation does not apply to SSM and diffusion models because these models do not expose its assumed standard attention–FFN block structure. PD disaggregation does not apply because diffusion models lack a prefill–decode separation.} In contrast, \name is model-agnostic and applies uniformly across diverse model architectures, providing strong generality.

\MyPara{Cost efficiency.}
Table~\textcolor{DarkGreen}{\ref{tab:cost_eff}} reports cost efficiency (Perf/\$), computed as the average throughput across five workloads divided by GPU rental cost and normalized to the left-hand homogeneous-GPU baseline. GPU rental prices are listed in Table~\textcolor{DarkGreen}{\ref{tab:GPU heterogeneity}}, and for disaggregation approachs, the total cost is the sum of both GPUs. Across all three heterogeneous GPU pairs, \name consistently achieves the highest cost efficiency, outperforming PD disaggregation by 1.5$\times$ on average (up to 1.6$\times$) and AF disaggregation by 1.4$\times$ on average (up to 1.5$\times$).

\modify{The H100+RTX Pro 6000 pair provides a concrete example of the benefits of heterogeneous execution. Across the five offline workloads, \name averages a 2.3$\times$ throughput speedup over a single H100 and exceeds twice the measured single-H100 throughput on four workloads. Under the rental-cost model in Table~\textcolor{DarkGreen}{\ref{tab:GPU heterogeneity}}, the heterogeneous pair has an aggregate rental cost 29.3\% lower than that of two H100s and achieves 1.64$\times$ the Perf/\$ of a single-H100 deployment. The 2.3$\times$ result compares throughput, whereas the 1.64$\times$ result compares throughput per dollar after accounting for aggregate GPU rental cost. The H100 favors bandwidth-sensitive kernels because of its higher memory bandwidth, whereas the RTX Pro 6000 benefits CUDA-core- or cache-sensitive kernels through its higher FP32 CUDA-core throughput and larger L2 cache. \name exploits these complementary strengths through kernel-level placement.}

These results suggest that cloud providers can improve cost efficiency by pairing high-end GPUs with more affordable counterparts, rather than provisioning additional expensive GPUs.

\begin{table}[t]
\centering
\footnotesize
\caption{Cost efficiency (Perf/\$) on three GPU pairs, normalized to the homogeneous left GPU baseline.}
\label{tab:cost_eff}
\begin{tabular}{cccc}
\hline
 & A100+L40s & H100+RTX Pro 6000 & B200+H100 \\
\hline
Homo.\ (left)   & 1.00 & 1.00 & 1.00 \\
Homo.\ (right)  & 1.18 & 2.01 & 0.78 \\
PD Dis.          & 0.87 & 1.00 & 0.70 \\
AF Dis.          & 1.02 & 1.07 & 0.74 \\
\name            & \textbf{1.21} & \textbf{1.64} & \textbf{1.01} \\
\hline
\end{tabular}
\end{table}

\begin{figure}[t]
    \centering
    \includegraphics[width=\linewidth]{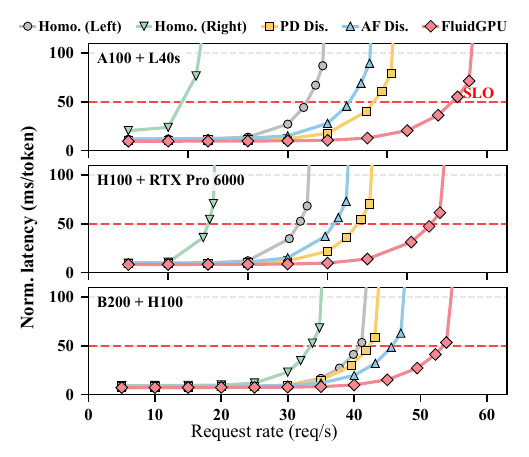}
    \caption{Online normalized latency on GPT-oss 20B across three GPU pairs. The red dashed line marks the SLO target (50 ms/token).}
    \label{fig:online_latency}
\end{figure}

\MyPara{Online latency.}
We evaluate GPT-oss 20B serving with Poisson arrivals at varying request rates. We measure normalized latency as the end-to-end request latency divided by output length in tokens. Figure~\textcolor{DarkGreen}{\ref{fig:online_latency}} reports normalized latency across different request rates on three GPU pairs. At low request rates, PD disaggregation and AF disaggregation have normalized latencies 1.3$\times$ and 1.2$\times$ higher than \name, respectively, because \name's latency-oriented policy minimizes per-request critical-path delay. Moreover, under an SLO target of 50 ms/token, \name sustains up to 1.3$\times$ higher request rate than the best baseline before violating the constraint.

\begin{figure}[t]
    \centering
    \includegraphics[width=0.98\linewidth]{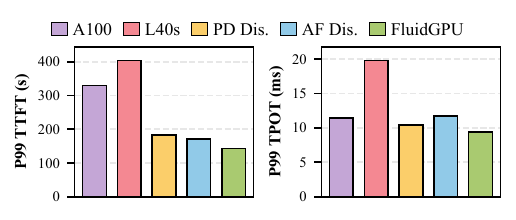}
    \caption{\modify{(a) P99 TTFT and (b) P99 TPOT for GPT-oss 20B under 16 req/s on A100+L40s.}}
    \label{fig:ttft_tpot}
\end{figure}

\modify{Figure~\textcolor{DarkGreen}{\ref{fig:ttft_tpot}} reports the P99 time-to-first-token (TTFT) and P99 time-per-output-token (TPOT) for GPT-oss 20B under 16 req/s on A100+L40s. \name achieves the lowest tail latency for both metrics. Its higher throughput lets the system process queued requests more quickly, reducing queueing delay and therefore P99 TTFT. Its kernel-level disaggregation places each kernel on the GPU best suited to its resource characteristics, shortening kernel execution along each request's critical path and therefore P99 TPOT. PD disaggregation has 1.28$\times$ higher P99 TTFT and 1.11$\times$ higher P99 TPOT than \name. AF disaggregation has 1.20$\times$ higher P99 TTFT and 1.26$\times$ higher P99 TPOT than \name.}

Compared to the offline setting, kernel disaggregation faces additional challenges in online serving. Inter-GPU transfer latency lies on the critical path of individual requests, and at low request rates, the pipeline cannot be fully saturated, limiting opportunities to overlap communication with computation. Nevertheless, \name still achieves the lowest latency among all disaggregation baselines, enabled by its latency-oriented policy and the queueing-aware policy switching of the online monitor.

\subsection{Cluster-Scale Results}
\label{sec:eval_cluster}

\begin{figure}[t]
    \centering
    \includegraphics[width=0.96\linewidth]{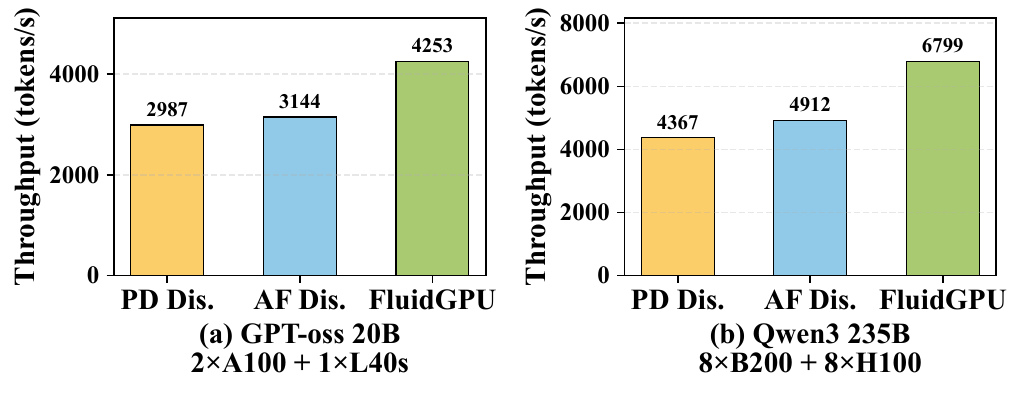}
    \caption{Cluster-scale offline throughput. (a) GPT-oss 20B on 2 A100 + 1 L40s. \modify{(b) Qwen3 235B on 8 B200 + 8 H100.}}
    \label{fig:cluster_scale}
\end{figure}

The preceding experiments evaluate \name on individual heterogeneous GPU pairs. \modify{We now examine its scalability on asymmetric GPU clusters and its composability with model parallelism (\eg, TP) in two offline setups. The first runs GPT-oss 20B on 2 A100 + 1 L40s, and the second runs Qwen3 235B~\cite{qwen3235B} on 8 B200 + 8 H100.} For the asymmetric 2 A100 + 1 L40s setup, \name's policy planner directly formulates a 3-GPU MILP, jointly optimizing kernel placement across all three GPUs. PD disaggregation uses L40s as the prefill instance and each A100 as an independent decode instance, and AF disaggregation runs attention on 2 A100 with TP=2 and FFN on L40s. Both baselines follow the default configurations from their original work~\cite{distserve, megascale-infer} and are tuned for maximum throughput. For the 8 B200+8 H100 setup, \name treats each B200+H100 pair as one TP rank and applies kernel disaggregation within each pair. In each setup, TP runs over NVLink within the homogeneous high-end GPU group, and collectives remain on the original GPU group.

Figure~\textcolor{DarkGreen}{\ref{fig:cluster_scale}} reports the offline throughput. Across both setups, \name achieves 1.5$\times$ the throughput of PD disaggregation and 1.4$\times$ that of AF disaggregation. The fundamental limitations of coarse-grained disaggregation persist at the cluster scale. In contrast, \modify{\name either applies kernel disaggregation independently within each TP rank pair or jointly optimizes kernel placement across the entire cluster while reusing the same policy planner and pipelined execution.} These results confirm that \name composes naturally with model parallelism and scales from a single GPU pair to multi-GPU clusters without requiring changes to the parallelism strategy.

\subsection{Performance Drilldown}
\label{sec:eval_ablation}

\begin{figure}[t]
    \centering
    \begin{minipage}[t]{0.57\linewidth}
        \vspace{0pt}
        \includegraphics[width=\linewidth]{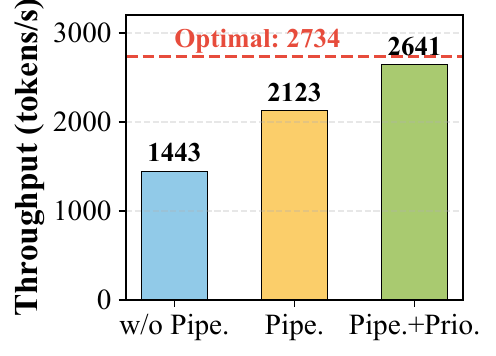}
    \end{minipage}\hfill
    \begin{minipage}[t]{0.4\linewidth}
        \vspace{0pt}
        \includegraphics[width=\linewidth]{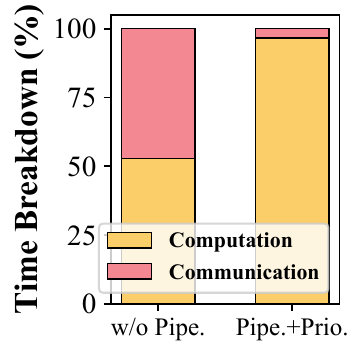}
    \end{minipage}
    \caption{(a) Throughput under different pipeline configurations. (b) Time breakdown on the bottleneck GPU.}
    \label{fig:pipeline_effective}
\end{figure}

\MyPara{Pipeline effectiveness.}
We evaluate pipelined request processing on GPT-oss 20B in an offline setting. Figure~\textcolor{DarkGreen}{\ref{fig:pipeline_effective}(a)} reports throughput under three configurations. The red dashed line denotes the optimal throughput, assuming zero communication overhead. Without pipelining, GPUs idle during cross-GPU transfers, achieving only about half of the optimal throughput. Pipelining without priority scheduling improves throughput by 1.47$\times$ by overlapping communication with computation from concurrent requests. Incorporating priority-aware scheduling further increases throughput to 96.6\% of optimal. Without priority differentiation, concurrent requests may compete for SMs and reach their communication phases simultaneously, leaving the GPU idle while all requests wait for data transfers. By assigning lower stream priorities to later-arriving requests, the GPU hardware scheduler preferentially allocates SMs to earlier requests, staggering their communication phases and keeping the GPU continuously occupied.

Figure~\textcolor{DarkGreen}{\ref{fig:pipeline_effective}(b)} shows the time breakdown on the bottleneck GPU. With pipelining and priority scheduling, communication idle time falls from nearly 50\% to less than 4\%, confirming that the pipeline effectively hides transfer latency. This result also validates the cost model in \S\textcolor{DarkGreen}{\ref{sec:policy_planner}}, which assumes that communication can be fully overlapped under steady-state pipelining.

\begin{figure}[t]
    \centering
    \includegraphics[width=0.97\linewidth]{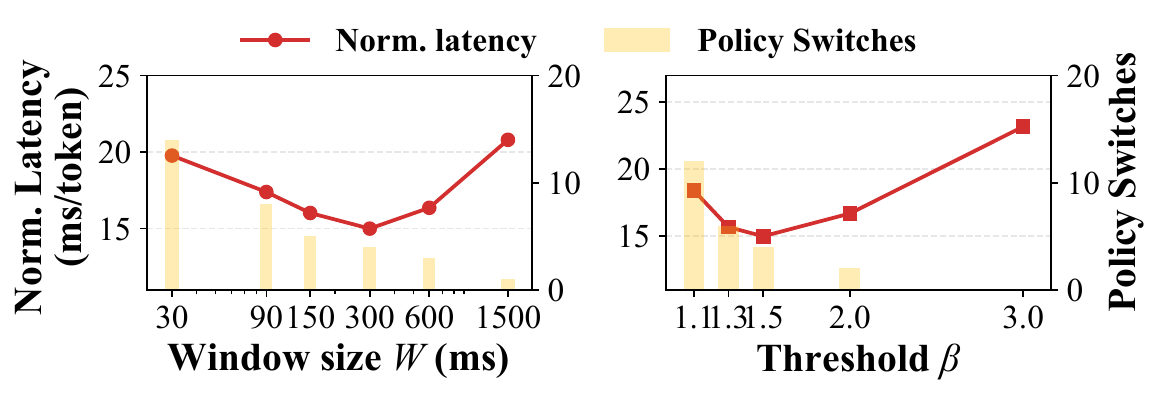}
    \caption{Sensitivity of normalized latency and policy switch count to window size $W$ ($\beta=1.5$) and queueing threshold $\beta$ ($W=300$ ms) on GPT-oss 20B.}
    \label{fig:sensitivity}
\end{figure}

\MyPara{Online monitor sensitivity.}
\label{sec:eval_online_monitor}
Figure~\textcolor{DarkGreen}{\ref{fig:sensitivity}} illustrates the sensitivity of normalized latency and policy-switch frequency to the window size $W$ and queueing threshold $\beta$ on GPT-oss 20B under a 360-second real-world workload trace~\cite{splitwise}. Within each window $W$, the monitor decides whether to trigger a policy switch based on $\beta$. The key trade-off lies between switching overhead and adaptation delay.
Each policy switch requires synchronizing all GPU workers at an iteration boundary, incurring a stall of approximately 30 ms. Aggressive configurations (\eg, $W=30$ ms or $\beta=1.1$) cause frequent switching, whose accumulated overhead increases latency by a factor of up to 1.32$\times$. In contrast, overly conservative settings (\eg, $W=1.5$ s or $\beta=3.0$) delay adaptation to load changes, leaving the system under suboptimal policies during spikes and increasing latency by a factor of up to 1.55$\times$. We therefore use $W=300$ ms and $\beta=1.5$ by default, which provide the best trade-off between responsiveness and overhead.

\begin{figure}[t]
    \centering
    \includegraphics[width=0.54\linewidth]{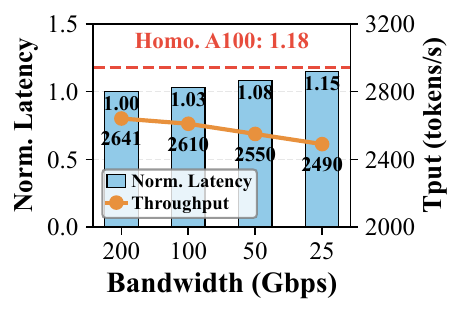}\hfill
    \includegraphics[width=0.44\linewidth]{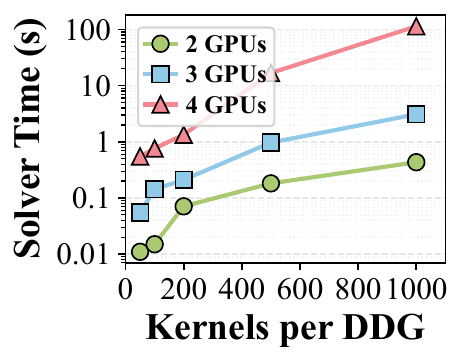}
    \caption{(a) Impact of network bandwidth on GPT-oss 20B with A100+L40s. (b) MILP solver time.}
    \label{fig:slow_network}
\end{figure}

\MyPara{Robustness to reduced bandwidth.}
We throttle the A100--L40s interconnect from 200 Gbps to 100, 50, and 25 Gbps and evaluate GPT-oss 20B in offline serving and in online serving under light load, as shown in Figure~\textcolor{DarkGreen}{\ref{fig:slow_network}}(a). For offline serving, even at 25 Gbps, \name's throughput drops by less than 6\% because pipelined request processing hides the increased transfer latency behind computation from concurrent requests. For online serving under a light request load, \name's latency-oriented policy accounts for the higher communication cost and shifts more kernels to the same GPU to reduce cross-GPU data transfer. Even with a 25 Gbps network, \name's normalized latency remains lower than the homogeneous A100 baseline, because the latency-oriented policy planner still finds a beneficial kernel placement at this bandwidth. In the worst case, \name keeps all kernels on A100, avoids cross-GPU transfers, and falls back to homogeneous-GPU execution without a performance cliff.

\MyPara{\modify{Offline preprocessing overhead.}}
\modify{In our evaluation, PTX instrumentation and dependency extraction take no more than 10 seconds during preprocessing. This measurement covers PTX rewriting, instrumented execution, and RAW-dependency extraction, but excludes per-GPU kernel latency profiling and policy optimization. By contrast, the approximately 5-second cost reported in \S\textcolor{DarkGreen}{\ref{sec:policy_planner}} measures MILP optimization over all profiled execution patterns after their DDGs and kernel profiles are available. Cached DDGs and policies are reused across requests, and runtime execution performs no instruction-level or element-level dependency tracking.}

\MyPara{MILP scalability.}
Figure~\textcolor{DarkGreen}{\ref{fig:slow_network}}(b) reports the MILP solver time as a function of kernel count and GPU count using Gurobi~\cite{Gurobi}. For the 2-GPU setting, solving a single DDG takes 0.07 s for 200 kernels and 0.43 s for 1000 kernels. \modify{Using DeepSeek-V3 671B~\cite{deepseekv3} as a stress test, a forward iteration with one DDG containing nearly 1500 kernels can be solved within 1 s. Although solver time increases with the number of GPUs, most layers in large models share identical kernel compositions, enabling the same placement to be reused across repeated layers.} \name exploits this redundancy to reduce the problem size and further accelerate the MILP solver.

\section{Related Work}

\MyPara{Disaggregated inference serving.}
Recent work has explored disaggregation to better align workload characteristics with hardware features. DistServe~\cite{distserve} and Splitwise~\cite{splitwise} separate the prefill and decode phases of LLM inference across different GPU pools. \modify{HexGen-2~\cite{hexgen}} formulates the placement of disaggregated prefill and decode computations on heterogeneous GPUs as a constrained optimization problem, jointly considering resource allocation and parallelization strategies to improve serving throughput. Cronus~\cite{cronus} further enables fine-grained prefill partitioning to mitigate load imbalance arising from heterogeneous GPU capabilities. MegaScale-Infer~\cite{megascale-infer} disaggregates execution at the block level by separating attention and FFN components across GPUs with different hardware strengths.
\modify{These approaches nevertheless operate at phase or block granularity and remain tightly coupled to specific model architectures. In contrast, \name disaggregates at the kernel level, providing finer-grained optimization while remaining applicable across diverse model types.}

\MyPara{\modify{Cluster-level placement and scheduling.}}
\modify{Beyond choosing where to disaggregate computation within an inference request, another line of work optimizes model placement, request assignment, and parallelism configurations at the cluster level. Sailor~\cite{sailor} jointly searches data, tensor, and pipeline parallelism configurations for distributed training. Metis~\cite{Metis} uses workload-aware partitioning to automate training on heterogeneous GPUs. Hetis~\cite{hetis} combines selective operator parallelization with online request dispatching to balance serving load. Helix~\cite{helix} jointly optimizes model placement and request routing by representing heterogeneous GPU and network capacities through a max-flow abstraction.}

\modify{Among these systems, Helix is the closest to \name because both make heterogeneity-aware placement decisions. Their optimization scopes differ. Helix usually assigns contiguous layer groups to GPU nodes and routes requests through the resulting pipelines. \name uses this cluster-level placement and routing as fixed input, and then maps individual CUDA kernels within each heterogeneous GPU group using measured kernel latency and communication cost. Since Transformer layers largely repeat the same kernel composition, layer-level placement cannot exploit different hardware preferences among kernels within a layer. The two systems are therefore complementary rather than fully orthogonal. Helix can first determine cluster-level layer placement and request routes, while its layer-group boundaries become fixed communication boundaries for \name. Helix requires less fine-grained profiling and creates fewer placement boundaries. \name incurs one-time dependency-analysis and per-kernel profiling costs and models additional transfers in exchange for finer hardware--kernel alignment.}

\MyPara{\modify{Heterogeneous accelerator scheduling.}}
\modify{General-purpose runtimes including StarPU~\cite{starpu}, OmpSs~\cite{ompss}, Legion~\cite{legion}, and PaRSEC~\cite{parsec} schedule application tasks across heterogeneous processors and manage the associated data movement. These systems typically require applications to expose work and dependencies through runtime-specific abstractions such as task graphs. \name targets a different integration point. It starts from CUDA kernels already emitted by existing AI inference frameworks, automatically extracts their memory dependencies, and performs communication-aware kernel placement across heterogeneous GPUs.}

\section{Conclusion}
\modify{This paper has presented \name, the first runtime system that exploits heterogeneous GPUs through fine-grained kernel disaggregation. Our central finding is that kernels within the same inference workload have distinct hardware preferences, making the kernel a natural granularity for aligning workload heterogeneity with GPU architectural diversity. \name combines automatic PTX-level dependency extraction, communication-aware policy planning, and pipelined execution to improve throughput and cost efficiency across diverse model architectures.}

\section*{Acknowledgments}
We thank the anonymous reviewers for their comments and are particularly grateful to our shepherd for valuable feedback. This work was supported in part by the National Key Research and Development Program of China (No.~2024YFE0204100), the Pioneer Initiative Talent Program B of the Chinese Academy of Sciences (Nos.~E545030 and E401430), the Fundamental and Interdisciplinary Disciplines Breakthrough Plan of the Ministry of Education of China (No.~JYB2025XDXM118), and the National Natural Science Foundation of China under Grant Nos.~U25A6023, 92464301, 62302479, U23B2020, and 625B2002. Tao Xie is also affiliated with the Key Laboratory of High Confidence Software Technologies (Peking University), Ministry of Education, China; Beijing Tongming Lake Information Technology Application Innovation Center, China; Institute of Systems for Advanced Computing at Fudan University, China; Shanghai Research Institute of Systems of Open Computing, China; Southeast University, Nanjing, China. Tiancheng Hu, Junhao Hu, and Yuzheng Wang are also affiliated with the Key Laboratory of High Confidence Software Technologies (Peking University).
Ting Cao is affiliated with the Institute for AI Industry Research, Tsinghua University, Beijing, China.
Corresponding authors: Tao Xie and Chenxi Wang.

\IEEEtriggeratref{62}
\bibliographystyle{IEEEtran}
\bibliography{reference}
\end{document}